\documentstyle[prb,aps,epsfig]{revtex}

\begin{document}

\draft

 
\title{ Magnons in the half-doped manganites }

\author {C. I. Ventura and B. Alascio}

\address {Centro At\'omico Bariloche, 8400-Bariloche, Argentina.}

\maketitle

\begin{abstract}

{\small 

Recently, based on the refined crystal structure of 
Pr$_{0.6}$Ca$_{0.4}$MnO$_{3}$ from neutron diffraction, 
Daoud-Aladine et al.[Phys.Rev.Lett.{\bf 89},97205(2002)] 
 have proposed a new ground state structure 
for the half-doped manganites R$_{0.5}$Ca$_{0.5}$MnO$_3$, where R is a  
trivalent ion like Bi, La, Pr, Sm or Y. 
Their proposal describes the CE magnetic structure attributed to   
these materials as an arrangement of dimers along the    
ferromagnetic Mn zig-zag chains that form it. The formation of dimers
is in accordance with previous theoretical calculations  
 based on simplified models for these compounds.
However, the dimers' proposal is in conflict   
with the Goodenough-Kanamori-Anderson rules,   
which give a coherent description of many transition metal insulating   
compounds and predict the coexistence of Mn$^{3+}$ and Mn$^{4+}$ ions 
in equal parts in the half-doped manganites. On the other hand, 
Rivadulla et al. [Phys.Rev.B {\bf 66},174432(2002)] have studied several 
single crystal samples of half-doped manganites and 
propose a phase diagram in terms of the tolerance factor which  
contains both types of structures. 
In the present work we have calculated the magnon dispersion relations 
for the CE magnetic structure, arising  for each type of proposal: 
the charge ordered and the dimer phases, respectively. 
We consider a three-dimensional unit cell containing 16 spins, 
and compare the magnetic excitations along different paths in the 
first Brillouin zone. We conclude that measurement of the magnon
dispersion relations should allow a clear distinction between the 
two proposals, predicting qualitative differences arising 
along specific directions of propagation in the first Brillouin zone.

}   

\end{abstract}




\widetext

Half-doped mangani\-tes  of stoi\-chio\-me\-try  
  R$_{1-x}$Ca$_{x}$MnO$_{3}$, x= 1/2,  where R is a trivalent 
ion like Bi, La, Pr, Sm or Y,  display a rich phase diagram 
depending on their composition \cite{phdiags}: 
in the absence of  magnetic field, 
charge-ordered antiferromagnetic insulator phases (COA), 
ferromagnetic metallic (FM), antiferromagnetic insulator, and 
paramagnetic phases have been reported. Strong colossal 
magnetoresistance appears by driving the materials from COA to FM 
by application of moderate external magnetic fields.

Below the N\'eel temperature the half-doped compounds are found in the 
CE magnetic structure \cite{wollan}, 
in which ferromagnetic Mn zig-zag chains   
contained in parallel planes are found in an antiferromagnetic 
arrangement.     
 The description of the CE magnetic structure 
found in most half-doped manganites is based  on the well known 
Goodenough-Kanamori-Anderson (GKA) rules \cite{gka},  
implying a checkerboard ordering of Mn$^{3+}$/Mn$^{4+}$ ions in the
crystal \cite{good}, as shown in Fig.1. 


 ( +orbital)  ordered (CO) scenario for half-doped manganites in 


However, Daoud-Aladine et al.\cite{daoud} have shown recently that the 
crystal structures of Pr$_{0.60}$Ca$_{0.40}$MnO$_{3}$ single crystals   
obtained from neutron diffraction data can not be interpreted 
in terms of Mn$^{3+}$/Mn$^{4+}$ charge ordering.  
Rather, the data suggest an intermediate valence of 3.5 for all Mn 
ions. Furthermore, the XANES results on the Mn K-edge of  
La$_{0.50}$Ca$_{0.50}$MnO$_{3}$ by J.Garc\'{\i}a et al. \cite{jgarcia}
 suggest also the same
valence for Mn. On the basis of these results, Daoud-Aladine et
al. propose that the zig-zag ferromagnetic chains contained in the CE 
magnetic structure are formed by a succession of Mn dimers
(or Zener polarons) \cite{daoud}, such as indicated in Fig.2.
 Mn ions forming the dimers share one electron and point in the same
direction forming a large spin unit. The chains order
antiferromagnetically between them within the Mn structure. 
Theoretically, such situation had been studied  previously in one  
\cite{previoustheory1d}, two \cite{previoustheory2d} 
and three dimensional \cite{previoustheory3d} 
structures as a consequence of the competition  
between double exchange and superexchange. 




More recently, Rivadulla et al. \cite{rivadulla} have
proposed a new phase diagram for the half-doped manganites in terms of the
tolerance factor $t$. In the new phase diagram both phases appear: the
charge ordered phase is found for $t$ below a critical value
$t_{c}=0.975$, and the dimer phase for $t$ larger than $t_{c}$. Both
(insulating) phases are separated by a unique (metallic) ferromagnetic
phase.  Contrary to the usual metallic ferromagnetic phase of hole doped
manganites which is stabilized by pressure, the ferromagnetic phase that
separates the two insulating phases at half filling can be suppresed by
pressure. 

Since both insulating phases proposed for the CE magnetic structure, 
the charge ordered and the dimer one, have some common
characteristics like: their insulating character, both are
metamagnetic going into a metallic ferromagnetic phase under moderate
magnetic fields, include zig-zag ferromagnetic chains, etc., 
we decided to calculate their magnetic excitation spectra 
as a mean to distinguish between them, and also
as a way to check the difference of the magnetic interaction 
along the chains in the proposed models. 

Therefore, we have considered a basic three-dimensional unit cell 
containing 16 spins (Mn), half of them in each of two consecutive
planes, as shown in Fig. 3.
For the charge ordered phase, two different values of spin
(representing the $S_{1}=2$ and $S_{2}=3/2$ spins of Mn$^{3+}$ and Mn$^{4+}$,
respectively) would appear, in checkerboard arrangement like shown 
in Fig. 1. 
The interactions are superexchange couplings between all Mn ions:
one ferromagnetic coupling along nearest neighbour spins on a chain (F'=F),
and two antiferromagnetic nearest neighbour couplings: one inter-chain
coupling (A) and one inter-plane (A').   
On the other hand, for the dimer phase, only one value of
spin ($S_1=S_2$) would be present, while the ferromagnetic coupling between
nearest neighbour spins along a zig-zag chain 
now would take two different values: the interaction between the 
spins forming a dimer is double exchange (F'), while a superexchange
interaction (F) couples nearest neighbour spins in different dimers.




This state of things can be described by a model composed of a 
series of appropriate nearest neighbour Heisenberg coupling terms. 
With the notation for magnetic couplings introduced in Fig. 3, 
the following Hamiltonian has been considered: 
\begin{eqnarray}
H & = & - F'  \sum_{(n,\alpha; n',\alpha') / \in D}  {\bf S}_{n,\alpha} \cdot
{\bf S}_{n',\alpha'} 
- F  \sum_{(n,\alpha; n',\alpha') / NN, \in C, \not\in D}  {\bf
S}_{n,\alpha} \cdot
{\bf S}_{n',\alpha'} \nonumber \\
  & + & A  \sum_{(n,\alpha; n',\alpha') / NN, \not\in C, \in P} 
{\bf S}_{n,\alpha} \cdot {\bf S}_{n',\alpha'} 
+ A'  \sum_{(n,\alpha; n',\alpha') / NN, \not\in P} 
{\bf S}_{n,\alpha} \cdot {\bf S}_{n',\alpha'} \; .
\label{eq:Hreal}
\end{eqnarray}
We have identified the position of the spins by indices: $\alpha =(i,j,k)$ 
to locate the 3D unit cell in the Bravais lattice, 
and site labels $n (=1,..,16)$ to distinguish the spins inside one unit cell. 
Capital letters $D$ denote a dimer, $NN$ nearest neighbour sites, 
$C$ one chain, and $P$ one plane. $F,F',A$ and $A'$ are positive.  
For the charge-ordered phase the ferromagnetic couplings along a
zig-zag chain are F'=F, while each coupling term will include spins of
different magnitude (due to the checkerboard arrangement shown in
Fig. 1).
 
To calculate the magnetic excitations at low temperatures we have used 
the Holstein-Primakoff transformation in linear spin wave
approximation \cite{QTS?}, which in our case demanded the introduction
of 16 different kinds of boson operators, one for each spin inside the unit
cell of Fig. 3, distinguishing the spin-up and spin-down sublattices. 
That is, we introduced for the spin-up sublattice:
\begin{eqnarray} 
S_{n,\alpha}^{+} & = & \sqrt{2 S_n} \, a_{n,\alpha} \; \nonumber \\
S_{n,\alpha}^{-} & = & \sqrt{2 S_n} \, a_{n,\alpha}^{\dagger} \; \nonumber \\
S_{n,\alpha}^{z} & = & S_n - \, a_{n,\alpha}^{\dagger}a_{n,\alpha} \; , 
\end{eqnarray}
where $n = 1,2,3,4,13,14,15,16$ with the notation of Fig.3, and $S_n$
is the magnitude of spin $n$. Without loss of generality, we can put:  
$ S_n = S_{1}$, for $n=1,3,14,16$, and  $ S_n = S_{2}$, for
$n=2,4,13,15$ ( $S_1 = S_2$, for the dimer phase).

Similarly for the spin-down sublattice, we introduced:
\begin{eqnarray} 
S_{n,\alpha}^{+} & = & \sqrt{2 S_n} \, b_{n,\alpha}^{\dagger} \; \nonumber \\
S_{n,\alpha}^{-} & = & \sqrt{2 S_n} \, b_{n,\alpha} \; \nonumber \\
S_{n,\alpha}^{z} & = & - S_n + \, b_{n,\alpha}^{\dagger}b_{n,\alpha} \; , 
\end{eqnarray}
where $n = 5,6,7,8,9,10,11,12$ with the notation of Fig.3. 
Without loss of generality, we put:  
$ S_n = S_{1}$, for $n=5,7,10,12$, and  $ S_n = S_{2}$, for
$n=6,8,9,11$.

Introducing the Fourier transform of the boson operators (which for 
compactness we will denote here: $ a_{n,\vec{k}}^{(\dagger)} \equiv  
a_{n}^{(\dagger)} \; ; \;b_{n,\vec{k}}^{(\dagger)} \equiv  
b_{n}^{(\dagger)}$) , we then 
determined the Hamiltonian, 
which for the dimer phase ($ S_1 = S_2 \equiv S$) takes the 
following form:


\begin{eqnarray}
& H  =  \left[ F' + F + 2 ( A + A' ) \right] \; \left[ - 8 S^{2} N \;
  +  S    \; \left( \sum_{\vec{k}; n=1-4,13-16}a_{n}^{\dagger}
a_{n}  + 
 \sum_{\vec{k}; n=5-12}  b_{n}^{\dagger} b_{n}  \right) \right]
 \nonumber \\
 - & S F'  \sum_{ \vec{k} } \left[ a_{2}^{\dagger}a_{3} 
+ b_{5}^{\dagger}b_{6}
+  b_{10}^{\dagger}b_{11} +
 a_{13}^{\dagger}a_{14} + 
  e^{i k_x a_x} a_{1}^{\dagger} a_{4} +
e^{i k_y a_y} a_{15}^{\dagger} a_{16} +  
 e^{-i k_y a_y} b_{7}^{\dagger} b_{8} +
e^{-i k_x a_x} b_{9}^{\dagger} b_{12} + H.c. \right]
 \nonumber \\
- & S F  \sum_{ \vec{k} } \left[ a_{1}^{\dagger}a_{2} 
+ a_{3}^{\dagger}a_{4}
+  a_{14}^{\dagger}a_{15} + 
 b_{6}^{\dagger} b_{7}
+  b_{9}^{\dagger} b_{10} +
 b_{11}^{\dagger} b_{12} +
e^{-i (k_x a_x + k_y a_y) } b_{5}^{\dagger}b_{8} +
 e^{i (k_x a_x + k_y a_y)} a_{13}^{\dagger}a_{16} +
+ H.c. \right]
\nonumber \\
 + & S A   \sum_{\vec{k}} \left[ a_{1} b_{6} 
+ a_{3} b_{6}
+  a_{3} b_{8} +
 b_{7} a_{4} + 
  b_{9} a_{14} 
+ b_{11} a_{14}
+  b_{11} a_{16} +
 a_{15} b_{12} +
   e^{-i k_x a_x} a_{1}b_{8} +
e^{-i k_x a_x} a_{13}b_{12} + 
\right. \nonumber \\ & \left.
  e^{-i k_y a_y} a_{13}b_{10} +
e^{-i k_y a_y} a_{15}b_{10} + 
  e^{i k_x a_x} b_{5}a_{4} +
e^{i k_y a_y} b_{5}a_{2} + 
 e^{i k_y a_y} b_{7}a_{2} +
e^{i k_x a_x} b_{9}a_{16} + 
H.c. \right]
\nonumber \\
 + & S A'  \sum_{\vec{k}} \left[ (1+e^{-i k_z a_z})
 \left( a_{1}b_{9} 
+ a_{2}b_{10}
+  a_{3}b_{11} +
 a_{4}b_{12} \right) + 
 (1+e^{i k_z a_z}) \left(b_{5}a_{13} +
b_{6}a_{14} + 
  b_{7}a_{15} +
 b_{8}a_{16} \right) + H.c. \right] \; \; ,
\end{eqnarray}
where $N$ denotes the total number of 16-spin unit cells as in Fig.3.
 
A similar expression can be written for the Hamiltonian of the  
charge-ordered phase, taking into account the two values of spin  
which will appear ($S_1 \not= S_2$) and $F=F'$ for that case. 

The magnon excitations of the system for the dimer and the CO phases 
can then be obtained by diagonalization of the corresponding Hamiltonian.
Concretely, we can obtain them by diagonalization of the following 
16x16 general matrix (multiplied by $S \equiv \sqrt{S_1 S_2}$), 
which for appropriate values of its parameters (as mentioned above) 
describes the two phases as particular cases:  


\begin{equation}
{\scriptsize
\left[ 
\begin{array}{cccccccccccccccc}
\lambda_1 & -F & 0 & -F'\Phi_x & 0 & A & 0 & A\Phi_x 
& A'\theta_z & 0 & 0 & 0 & 0 & 0 & 0 & 0 \\ 
-F & \lambda_2 & -F' & 0 & A\Phi^{*}_y & 0 &  A\Phi^{*}_y & 0  
& 0 &  A'\theta_z &0  &0  &0  & 0 & 0 & 0 \\ 
0 & -F' & \lambda_1 &-F  &0  &A  & 0 &A 
& 0 & 0 & A'\theta_z  & 0 &0  &0  & 0 & 0 \\ 
-F'\Phi^{*}_x & 0 & -F & \lambda_2 & A\Phi^{*}_{x} & 0 & A & 0 
& 0 & 0 & 0 &  A'\theta_z &0  & 0 &0  & 0 \\ 
0 & -A\Phi_y & 0 & -A\Phi_x & -\lambda_1 & F' & 0 & F \Phi_x \Phi_y 
& 0 & 0 &0  &0  & - A'\theta_z & 0 &0  & 0 \\ 
-A & 0 & -A &0  &F'  &-\lambda_2  & F &0 
& 0 & 0 &0  &0  & 0 & - A'\theta_z &0  & 0 \\ 
0 & -A\Phi_y & 0 & -A & 0 & F & -\lambda_1 & F'\Phi_y
& 0 & 0 & 0 &0  &0  &0  &  -A'\theta_z & 0 \\ 
-A\Phi^{*}_x & 0 & -A & 0 & F\Phi^{*}_x\Phi^{*}_y & 0 &F'\Phi^{*}_y 
& -\lambda_2
&0  & 0 & 0 &0  & 0 & 0 &0  &  -A'\theta_z \\ 
-A'\theta^{*}_z & 0 & 0 & 0 & 0 & 0 & 0 &0 
& -\lambda_2 & F & 0 & F'\Phi_x & 0 & -A & 0 & -A\Phi_x \\ 
0 & - A'\theta^{*}_z  & 0 & 0 & 0 & 0 &0  &0 
& F & -\lambda_1 & F' & 0 & -A\Phi^{*}_y & 0 & -A\Phi^{*}_y & 0  \\ 
0 & 0 & -A'\theta^{*}_z & 0 &0  & 0 & 0 &0 
& 0 & F' & -\lambda_2 & F  &0  & -A  & 0 & -A  \\ 
0 & 0 & 0 & -A'\theta^{*}_z & 0 & 0 & 0 & 0
& F'\Phi^{*}_x & 0 & F & -\lambda_1 & -A\Phi^{*}_{x} & 0 & -A & 0 \\ 
0 & 0 & 0 &0  &  A'\theta^{*}_z & 0 &0  & 0
&0 & A\Phi_y & 0 & A\Phi_x & \lambda_2 & -F' & 0 & -F \Phi_x \Phi_y \\ 
0 & 0 & 0 & 0 & 0 & A'\theta^{*}_z  & 0  & 0
&A & 0 & A &0  &-F'  &\lambda_1  & -F &0  \\ 
0 & 0 &0  & 0 & 0 & 0 & A'\theta^{*}_z  & 0 
&0 & A\Phi_y & 0 & A & 0 & -F & \lambda_2 & -F'\Phi_y \\ 
0 & 0 & 0 & 0 & 0 & 0 & 0 &   A'\theta^{*}_z
& A\Phi^{*}_x & 0 & A & 0 & -F \Phi^{*}_x \Phi^{*}_y & 0 & -F'\Phi^{*}_y 
& \lambda_1 
\end{array}
\right]
}
\end{equation}
where the matrix columns (or rows) are related to the 16 boson
operators, arranged in increasing order of their site label, and   
we have used the following notation for the matrix elements: 
$ \lambda_1 = S_2 [ F + F'+ 2 (A+A') ]/S$; 
  $ \lambda_2 = S_1 [ F + F'+ 2 (A+A') ]/S$ ;
$\Phi_\alpha = \exp{(i k_{\alpha} a_{\alpha}) }$ , ($\alpha=x,y$) ; 
$ \theta_z = (1 + \exp{(i k_{z} a_{z})} ) $.   

In Fig. 4 we exhibit our result for the 16 magnon excitation branches 
of the charge ordered (CO) phase, plotted along 8 different paths
throughout the first Brillouin zone,  
using a typical set of relative values for the 
exchange coupling parameters for the system (in the present work we
take the interdimer exchange coupling $F$ as the energy unit, setting 
the overall energy scale)    
and a ratio of spin values  $S_1/S_2 = 1.3$, such as that for   
Mn$^{3+}$/Mn$^{4+}$ magnetic moments. The trajectories along 
the first Brillouin zone depicted extend between the following
points: $\Gamma  \equiv 0 = (0,0,0)$, $X=(\pi/a_x,0,0)$, $Y=(0,\pi/a_y,0)$,
 $Z=(0,0,\pi/a_z)$, $M=(\pi/a_x,\pi/a_y,0)$ and 
$L=(\pi/a_x ,\pi/a_y,\pi/a_z)$.





In Fig. 5 we exhibit the results calculated for the magnons 
corresponding to the dimer phase, along the same first Brillouin zone
trajectories. We have now considered: $S_1/S_2= 1$, $F'= 2 F$, 
and the rest of the parameters 
unchanged for the sake of simplifying the comparison with the CO magnons. 
We have verified that the same magnon excitations are obtained 
for this phase, if one chooses a different relative array of the
dimers on neighbour zig-zag chains.

Comparison of Figs. 4 and 5 reveals that a qualitative difference 
in the magnon branches predicted for the CO and the dimer phases 
can be clearly pointed out. Namely, that along the $\Gamma \longrightarrow Y$
and especially along the $Y \longrightarrow M$ paths 
in the first Brillouin zone
 8 CO magnon branches are doubly degenerate, 
while additional degeneracies are predicted for the respective 
magnons in the dimer phase. For $A'<A$ six distinct magnon branches
 (4 of them doubly degenerate, but other 2 with degeneration four)
appear along $\Gamma \longrightarrow Y$ (while for $A=A'$ as in Fig.5
there is a slight lifting of the extra degeneracy, being 8 branches observable)
, while 4 magnon branches (with degeneration four) appear along 
$Y \longrightarrow M$ for the dimer phase.
This splitting or lifting of the dimer phase degeneracies observable 
in the CO phase increases proportionally with the magnitude 
of the antiferromagnetic in-plane $A$ coupling 
(the effect disappears for $A=0$), 
being almost unaffected by the values of the other magnetic couplings.
  
The anisotropy exhibited by the magnon dispersion relations 
(of Figs.4 and 5) along $\Gamma \longrightarrow X$ with 
respect to those along $\Gamma \longrightarrow Y$ 
can be understood considering the spin ordering along the x and y 
directions 
(see Fig. 3). Along $x$, the direction of the zig-zag chains, 
 the system can be viewed as composed by consecutive chains: 
 one of parallel spins (e.g. up), 
followed by another with alternating spins, 
then again one of parallel spins (down), etc. While along $y$  
the system has more symmetry: it can be viewed as consecutive 
chains, all of them formed by alternating spins. 

Apart from the qualitative differences between the magnon spectra  
of the CO and dimer phases which were pointed out above  
(mostly related to the magnitude of the antiferromagnetic  
in-plane coupling $A$), 
there are also quantitative variations of the spectra related  
to the values of the other magnetic couplings considered.  
In particular, in Fig. 6 we exhibit the magnon dispersion curves 
for a set of parameters more suitable for the description 
of the dimer phase: 
with an intradimer ferromagnetic coupling ($F'$) an order of magnitude
larger than the interdimer ferromagnetic coupling ($F$). This 
corresponds to the large estimated values ($t \sim 100$ meV)  
for the effective hopping between Mn ions in these compounds 
\cite{jvalue,zheng}. Recent estimations of the ferromagnetic
interdimer exchange coupling $F$ in these compounds consider it to be
of the order of $10$ meV \cite{zheng}.   
  For strong intradimer ferromagnetic coupling 
as in Fig. 6, a large energy gap would then separate the magnons between 
higher (lying approximately around 0.1eV) and lower (around 0-0.01eV) 
energy modes, being the latter possibly the only ones 
easier to probe experimentally.   



Summarizing, we have calculated the magnetic excitations to be
expected for the charge ordered and the dimer phases which have been 
proposed for the CE magnetic structure of the half-doped manganites. 
We predict qualitative differences between the magnons related to
specific exchange couplings, 
which should allow 
to distinguish between these two phases in neutron scattering experiments. 
Not only are there lifted degeneracies in the CO phase along certain 
first Brillouin zone paths with respect to the dimer phase, but  
the presence of dimers with strong intradimer exchange coupling 
should result in a smaller number of magnons being detected by 
 the experiments at low energies, while a larger number of magnetic
excitations should be easily detected if the charge-ordered phase 
was the one being measured.

\begin{center}
{ACKNOWLEDGEMENTS}
\end{center}

We wish to acknowledge discussions with
J.Rodr\'{\i}guez-Carvajal. B.A. and C.I.V. are members 
of the Carrera del Investigador Cient\'{\i}fico of CO\-NI\-CET 
(Consejo Nacional de Investigaciones Cient\'{\i}ficas y T\'ecnicas, 
Argentina). B.A. is a researcher of C.N.E.A. (Comisi\'on Nacional de 
Energ\'{\i}a At\'omica, Argentina).

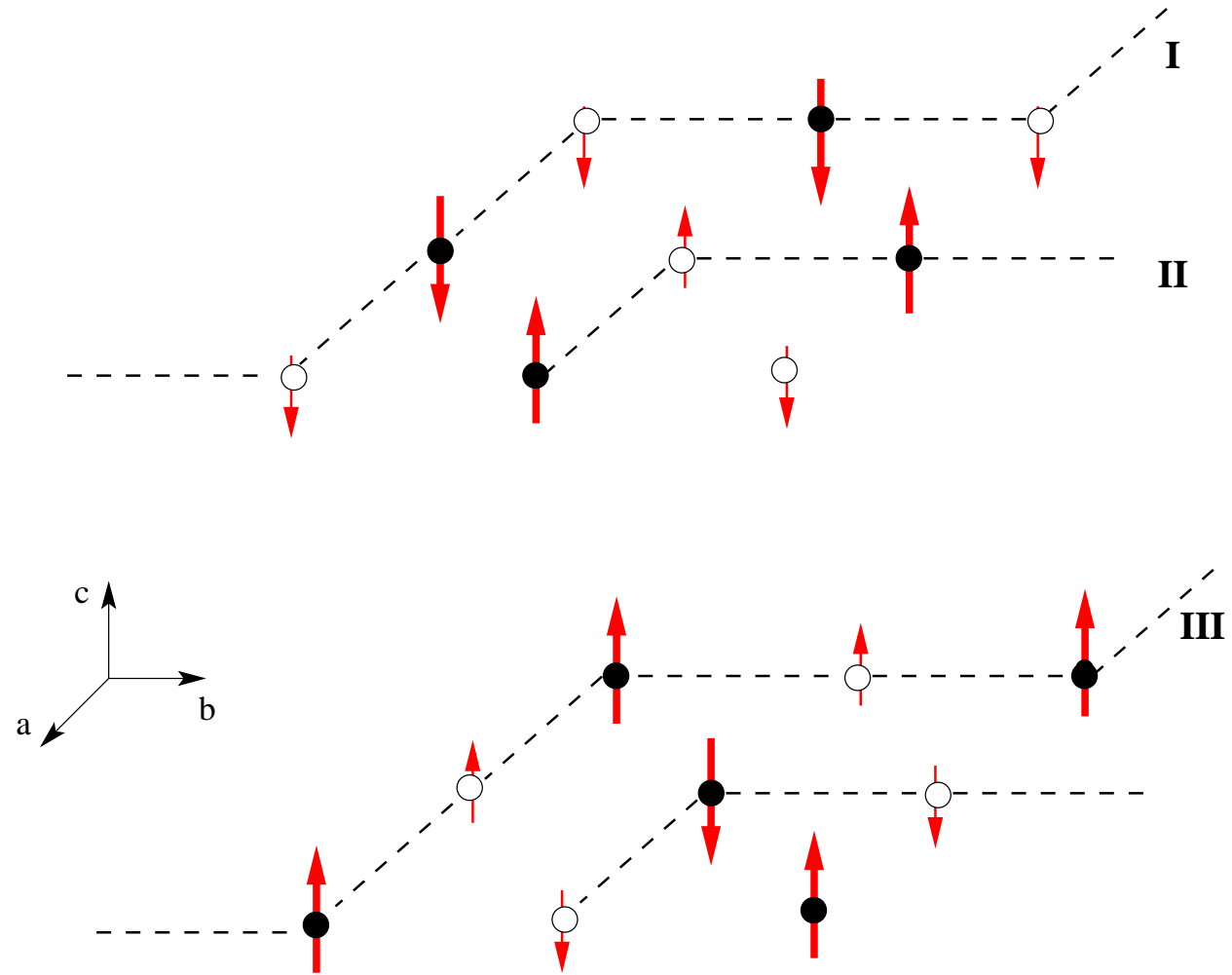
\begin{figure}
\caption{ Schematic picture of the Mn$^{3+}$/Mn$^{4+}$ charge
 ( +orbital)  ordered (CO) scenario for half-doped manganites in 
the CE magnetic structure. 
Large arrows denote $S=2$ ( Mn$^{3+}$) spins, while small 
arrows $S=1.5$ ( Mn$^{4+}$) spins.
I and II are coplanar ferromagnetic zig-zag chains, 
chain III is directly below I on a consecutive plane.}
\label{fig:co}
\end{figure}

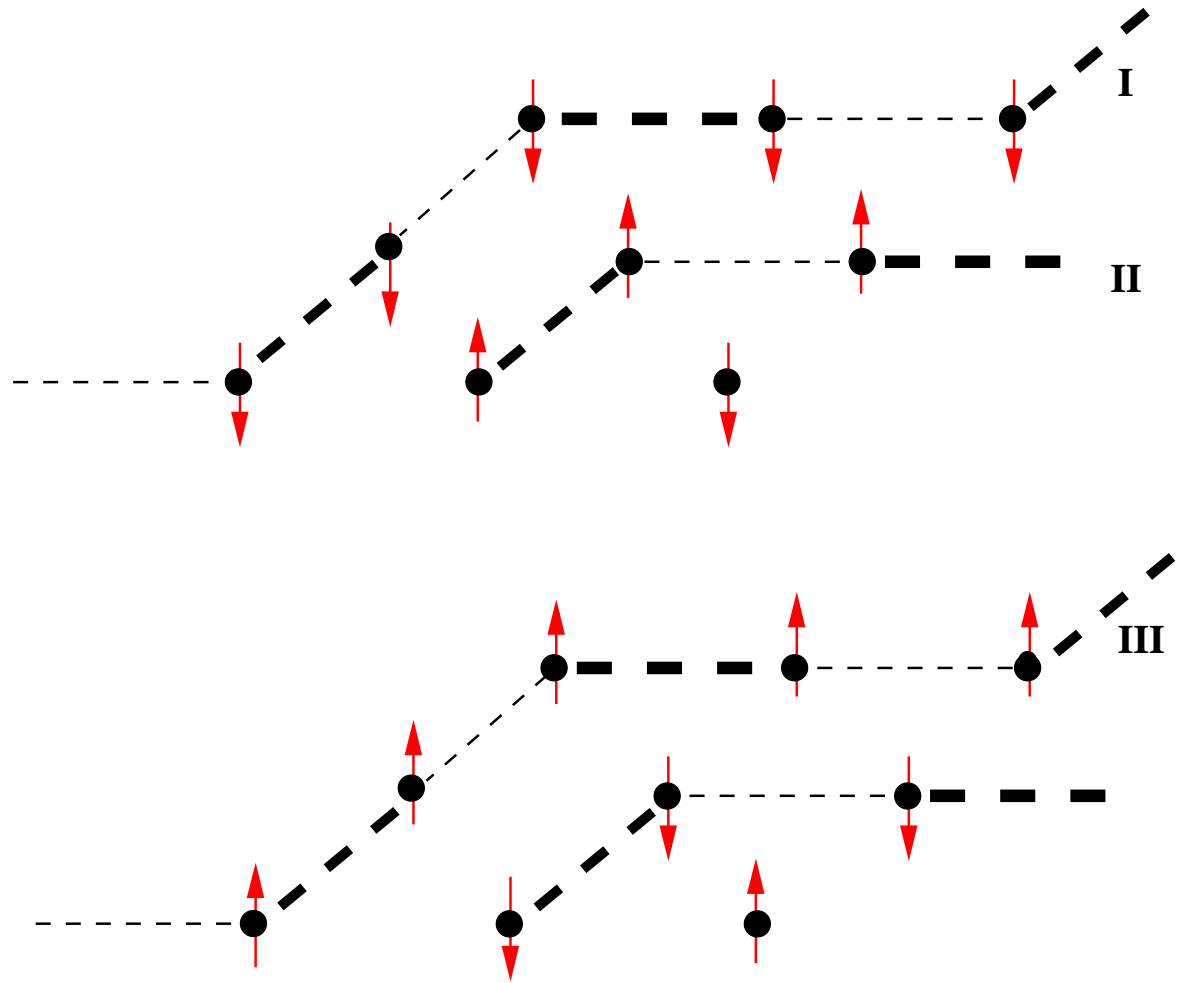
\begin{figure}
\caption{Schematic picture of the dimer phase for half-doped
manganites in the CE structure. 
All spins correspond to Mn ions with an intermediate valence of
$\sim$3.5. Thick dashed lines indicate dimers.}
\label{fig:dim}
\end{figure}

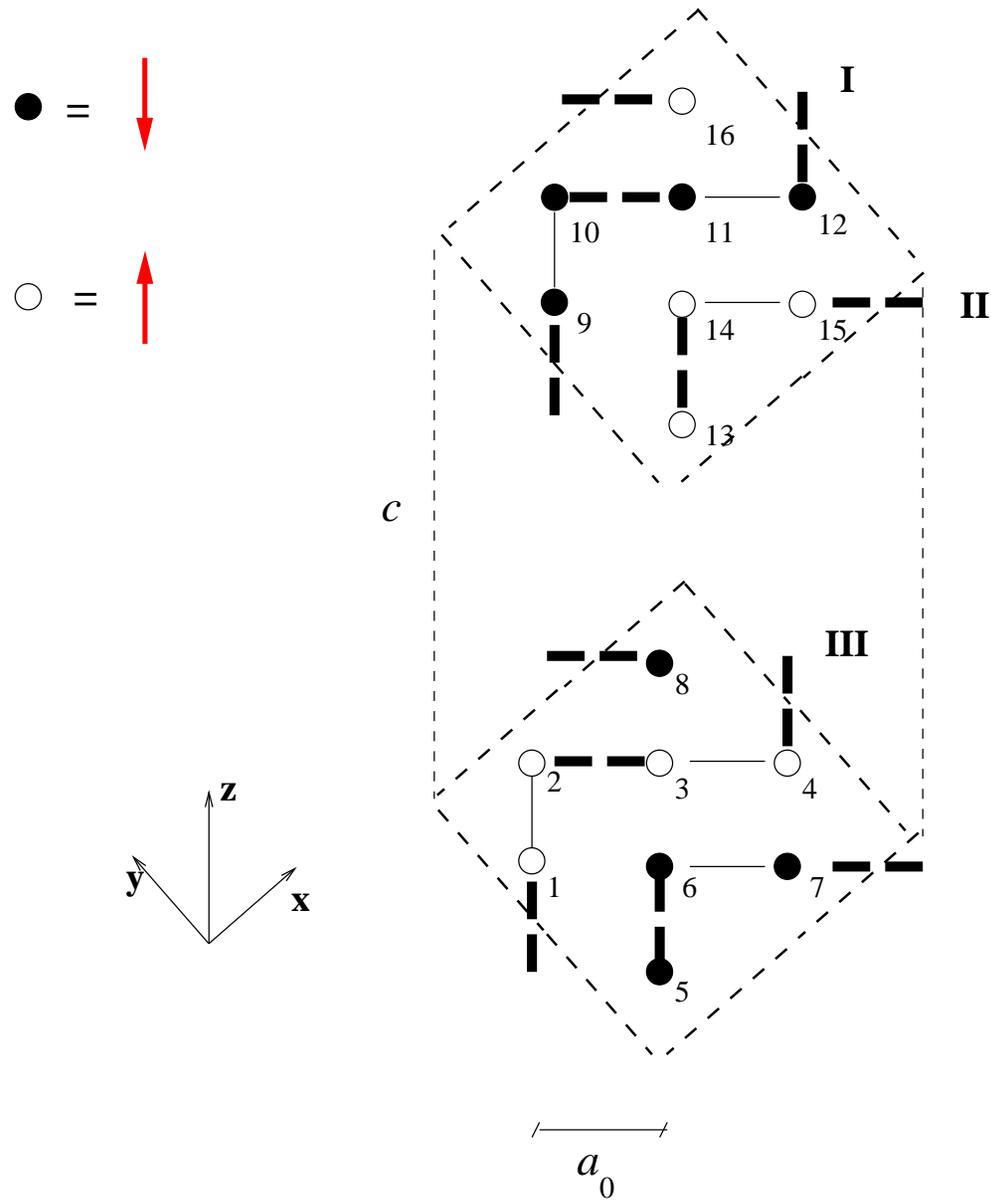
\begin{figure}
\caption{Schematic plot of the three-dimensional 16-spin unit cell
used ( Bravais lattice parameters: $ a_{x} = a_{y} = \sqrt{8} a_{0}$, 
$ a_{z} = 2 c $). 
Magnetic couplings between nearest neighbour spins: 
ferromagnetic ``intradimer'' F' (thick dashes) and ``interdimer'' F (solid
line) along the zig-zag chains (F'=F, in CO phase), 
antiferromagnetic interchain A for coplanar spins (e.g. spin 9 on
chain I and 14 on II) 
and  interplane A', for spins on chains in consecutive planes (e.g. 
spin 9 on chain I and 1 on III).}
\label{fig:unitcell}
\end{figure}


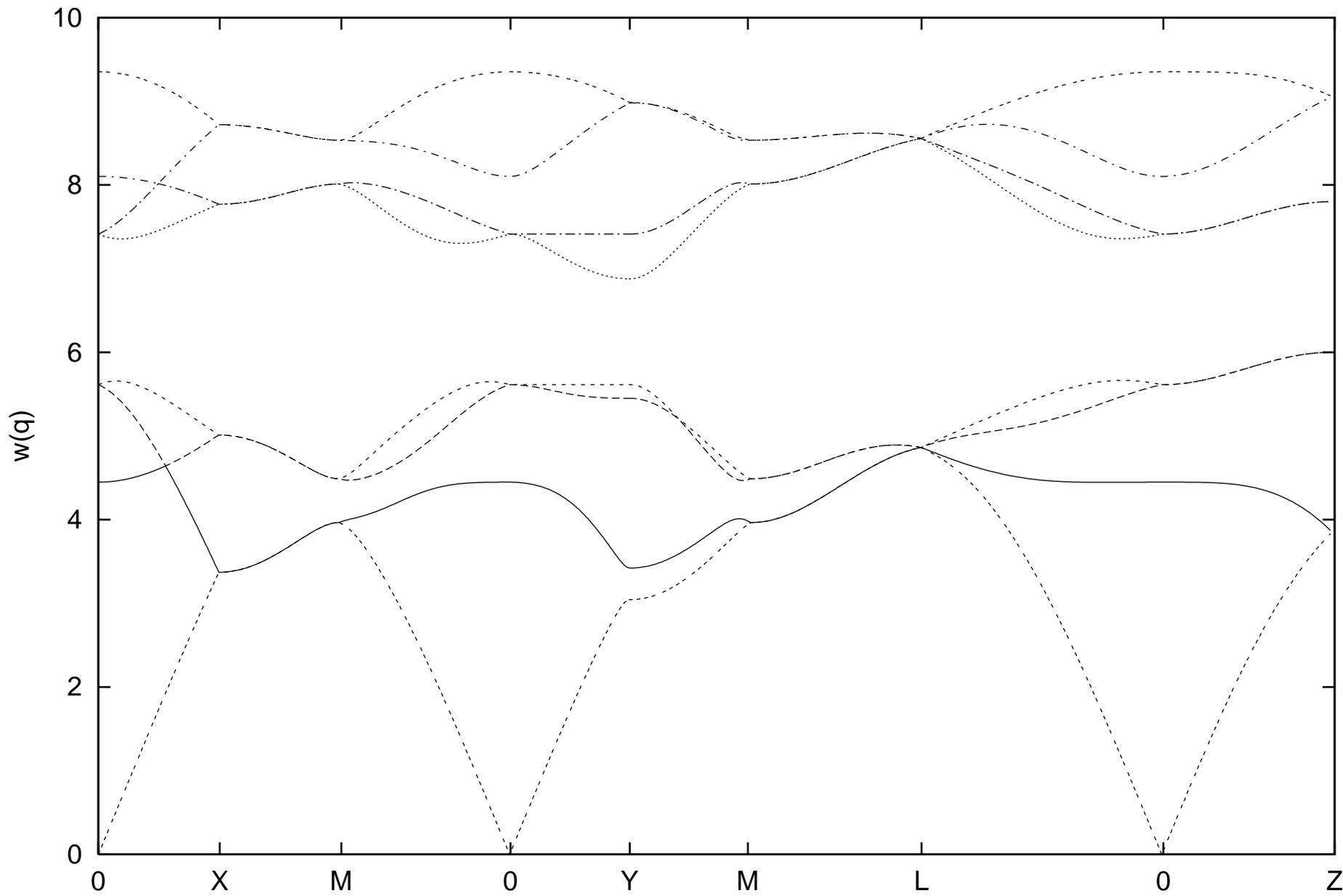
\begin{figure}
\caption{Magnetic excitations for the charge ordered (CO) phase, 
along 8 paths throughout the first Brillouin zone. $F$ is taken as 
the energy unit. Parameters:
$F'=F= 1$; $A= 1 =A'$; $S_{1} = 1.3 $, $S_{2}= 1$; $c/a_{0} = 1$. 
}
\label{fig:magco}
\end{figure}

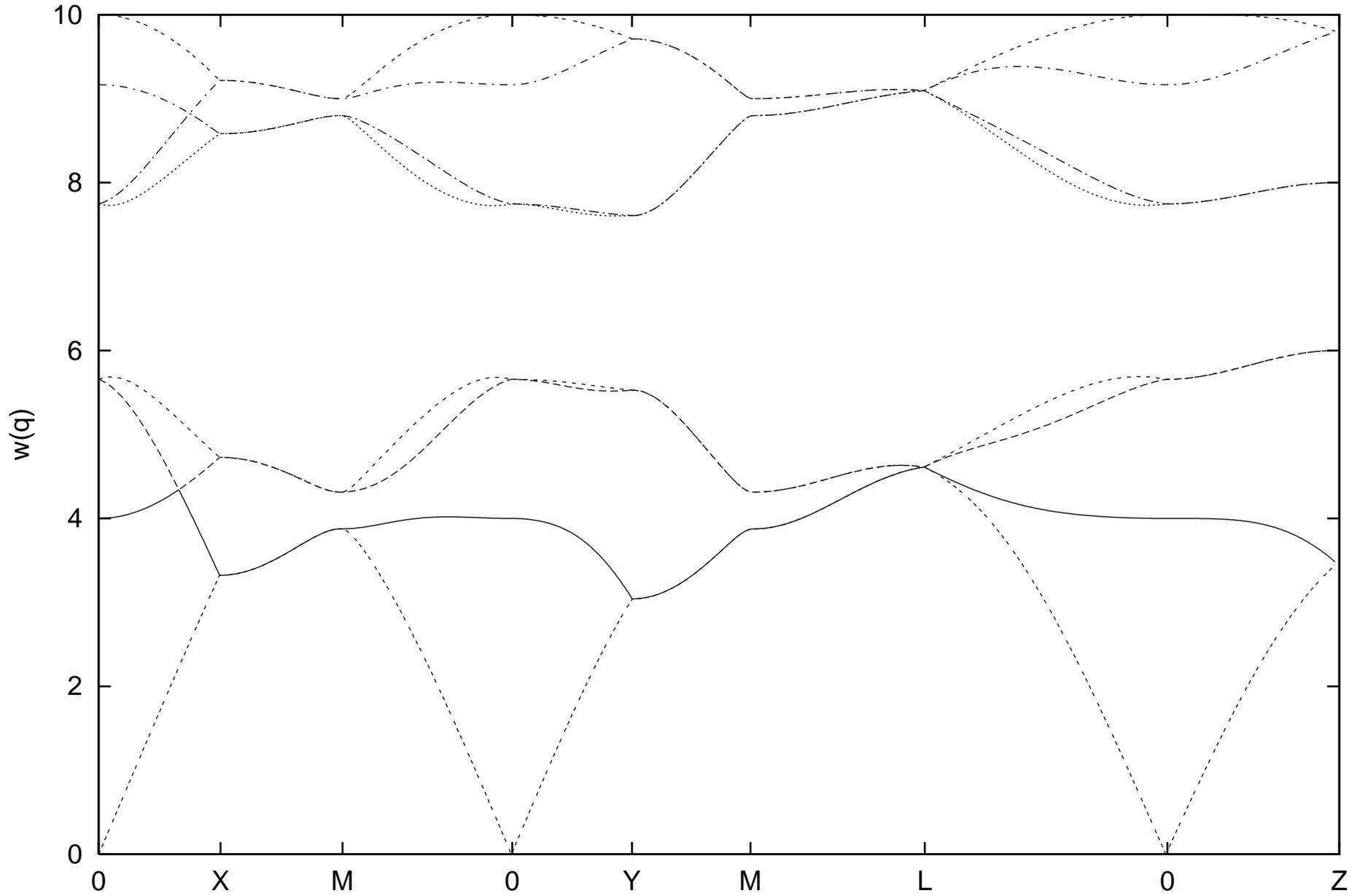
\begin{figure}
\caption{Magnetic excitations for the dimer phase, 
along 8 paths throughout the first Brillouin zone. Parameters:
$F'= 2 $, $F= 1$; $A= 1 =A'$; $S_{1}=S_{2}= 1$; $c/a_{0} = 1$. 
}
\label{fig:magdim}
\end{figure}

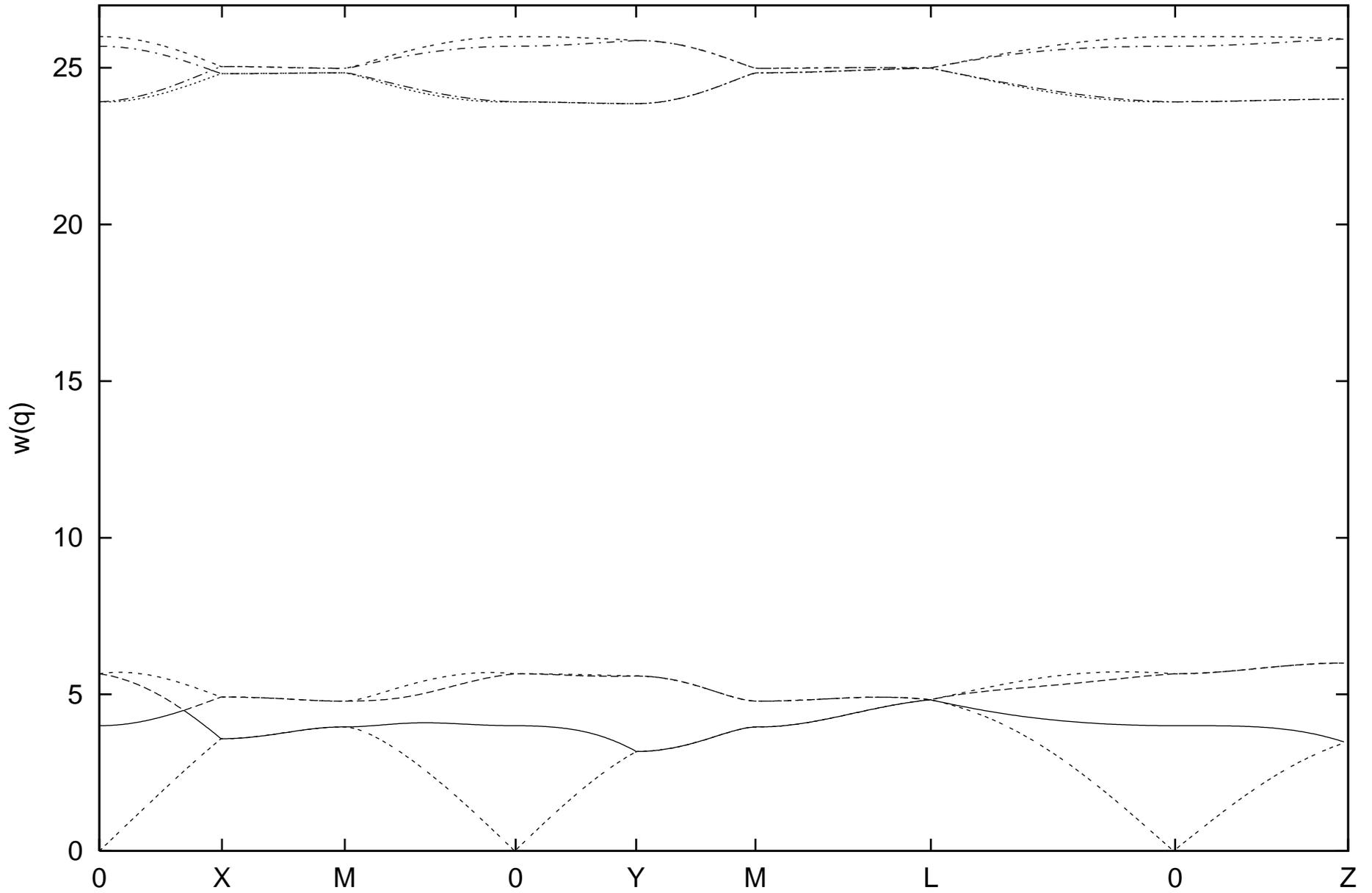
\begin{figure}
\caption{Magnetic excitations for the dimers phase, 
along 8 paths throughout the first Brillouin zone. For parameters:
$F'= 10 $, $F= 1$; $A= 1 =A'$; $S_{1}=S_{2}= 1$; $c/a_{0} = 1$. 
}
\label{fig:magdim2}
\end{figure}

\end{document}